\documentclass[aps,prb,showpacs,twocolumn]{revtex4-1}
\usepackage{amsfonts}
\usepackage{amssymb}
\usepackage{amsmath}
\usepackage{epsfig}
\usepackage{graphicx}
\usepackage{enumerate}
\usepackage{multirow}
\usepackage{verbatim}
\usepackage{color}

\begin{document}

\title{Low-noise conditional operation of singlet-triplet coupled quantum dot qubits}
\author{Shuo Yang}
\author{S. Das Sarma}
\affiliation{Condensed Matter Theory Center, Department of Physics,
University of Maryland, College Park, MD 20742}

\begin{abstract}
We theoretically study the influence of charge noise on a controlled phase gate, implemented using two proximal double quantum dots coupled electrostatically. Using the configuration interaction method, we present a full description of the conditional control scheme and quantitatively calculate the gate error arising from charge fluctuations. Our key finding is that the existence of noise-immune sweet spots depends on not only the energy detuning but also the device geometry. The conditions for sweet spots with minimal charge noise are predicted analytically and verified numerically. Going beyond the simple sweet-spot concept we demonstrate the existence of other optimal situations for fast and low-noise singlet-triplet two-qubit gates.
\end{abstract}


\maketitle

Coupled quantum dots are promising candidates for future implementations of quantum computation \cite{Wiel.03,Hanson.07,Loss.98,DiVincenzo.00,Levy.02,Laird.10,Petta.05}. They have potentially excellent scalability due to the well-developed semiconductor nanoelectronics technology. Moreover, the confinement potential can be electrically tuned by nearby lithographic gates, enabling easy controllability of quantum dynamics \cite{Petta.05,Pioro.08,Foletti.09,Barthel.10,Burkard.99,Hu.00}. However, such electrostatic controllability also makes the system vulnerable to electrical fluctuations in the environment \cite{Burkard.99,Hu.06,Culcer.09}, leading to decoherence, thus hindering the requisite coherent manipulation of quantum states. To minimize the influence of charge noise, one must search for optimal conditions or ``sweet spots'' in the parameter space \cite{Hu.06,Culcer.09,Stopa.08,Liqz.10,Nielsen.10}, where the first-order response of exchange energy to voltage fluctuations is zero, so that the decoherence of the exchange-coupled spin qubits to the environmental charge noise is minimal.

Recently, much attention has been focused on the multi-qubit coherent operations in coupled quantum dots \cite{Stepanenko.07,Shinkai.09,Petersson.09,Meunier.11,vanWeperen.11,Taylor.05,Taylor.07}. This is partly because all the needed single-qubit manipulations have already been demonstrated in experiments \cite{Pioro.08,Foletti.09,Barthel.10}, and partly because implementing elementary multi-qubit gates would be an important milestone en route to a scalable quantum computer. In a recent experiment, van Weperen \textit{et al.} \cite{vanWeperen.11} demonstrated the fast conditional operation of a singlet-triplet qubit controlled by an adjacent two-electron double quantum dot. A two-qubit controlled-phase (C-phase) gate is realized through the capacitive interaction between the double quantum dots (DQDs). However, the problem of charge fluctuations in this scheme is severe. It is therefore of vital importance to quantitatively understand charge noise and find multiqubit ``sweet spots'' for improving the performance and reliability of the two-qubit gate.

\begin{figure}
\includegraphics[width=0.95\columnwidth]{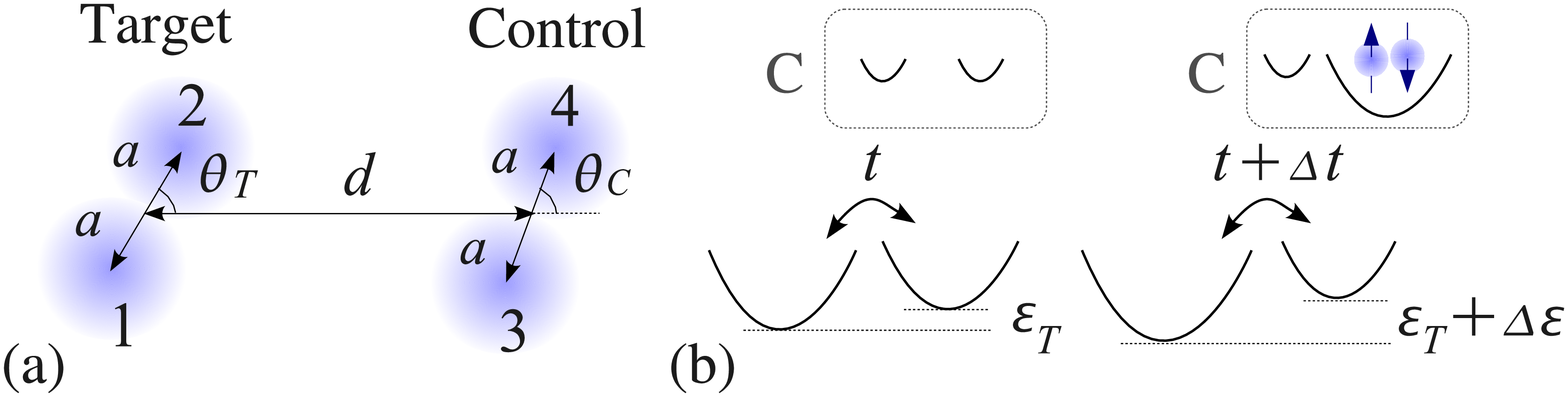}
\includegraphics[width=0.45\columnwidth]{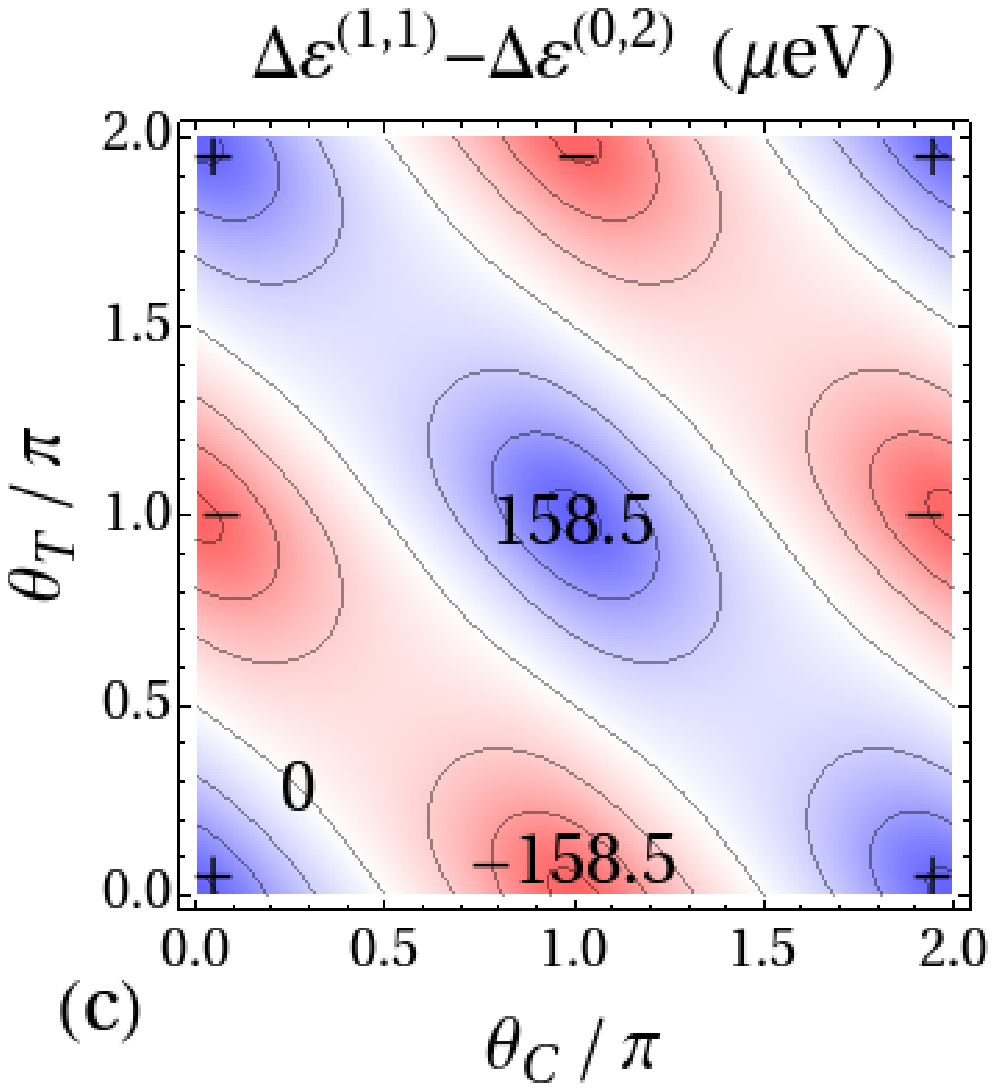}
\hspace{0.0\columnwidth}
\includegraphics[width=0.45\columnwidth]{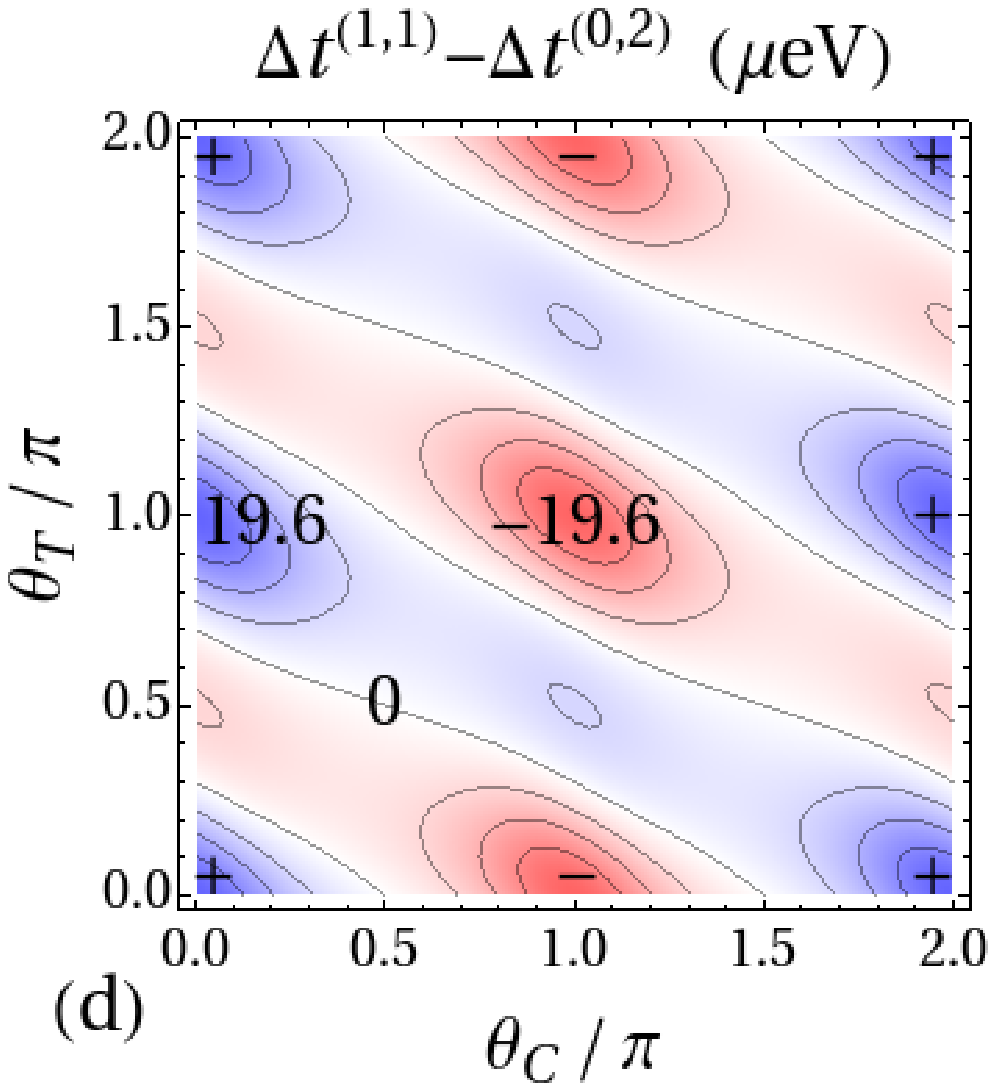}
\caption{(Color online) (a) General configuration of the control and target double quantum dots. (b) Loading electrons into the control DQD effectively changes the bias and tunnel coupling of the target DQD. (c)-(d) When the control DQD is tuned from $(0,2)$ to $(1,1)$ charge states, the variations of $\Delta \varepsilon$ and $\Delta t$ are plotted as functions of $\theta_{C}$ and $\theta_{T}$.}
\label{config}
\end{figure}

The goal of this theoretical work is to find sweet spots for the two-qubit conditional operation of the so-called singlet-triplet qubits \cite{Petta.05,Taylor.05,Taylor.07}, so that the impact of charge fluctuations is minimized. In particular, we explore for the first time the significance of the device geometry in reducing charge noise. In comparison to the previous investigations \cite{Stepanenko.07,Shinkai.09,Petersson.09,Meunier.11,vanWeperen.11,Taylor.05,Taylor.07}, a general arrangement of quantum dots with arbitrary geometry and relative configuration angles [Fig.~\ref{config}(a)] is studied in detail. Based on the multielectron configuration interaction calculations of the coupled double-dot system, we address the following questions: (i) How does the control mechanism depend on the device geometry? (ii) How is the C-phase gate affected by charge noise? (iii) Which configurations support sweet spots? (iv) What are the optimal detuning energies? (v) Are there any other favorable cases? The answers to these questions could be of considerable help to the ongoing double-dot multiqubit experiments in various laboratories worldwide.

We first introduce a microscopic model and provide a clear understanding of the conditional control scheme. In general, the Hamiltonian of a quantum dot system is given by $H=\sum_{kl} F_{kl} c_{k}^{\dagger} c_{l}$ $+\sum_{hjkl} G_{hjkl} c_{h}^{\dagger} c_{j}^{\dagger} c_{k} c_{l}$, where the sum includes all possible terms that conserve the total particle number and the total spin \cite{Yang.11,Wang.11}. In our case, quantum tunnel couplings between the control and target DQDs are highly suppressed because of the large potential barrier between the two qubits with each double-dot system being the usual singlet-triplet qubit, so that they are only capacitively coupled \cite{vanWeperen.11}. Hence, we have a simplified Hamiltonian $H=H_{C}+H_{T}+H_{{\rm int}}$.
Here $H_{C}$ and $H_{T}$ describe the isolated control and target DQDs \cite{Yang.11,Wang.11,JeJpJt}, and $H_{{\rm int}}$ describes their mutual interactions. More specifically, $H_{{\rm int}}$ includes three parts: the classical inter-dot Coulomb interactions $H_{U}=\sum_{ij} U_{ij} n_{i} n_{j}$, the occupation-modulated hoppings $H_{Jt}=-(J_{t1}n_{1}+J_{t2}n_{2})\sum_{\sigma}( c_{3 \sigma}^{\dagger} c_{4 \sigma} +{\rm H.c.})-(J_{t3}n_{3}+J_{t4}n_{4})\sum_{\sigma}( c_{1 \sigma}^{\dagger} c_{2 \sigma} +{\rm H.c.})$, and the hopping-hopping interactions $H_{Jh}=-J_{h} $ $\sum_{i,j,i',j'} c_{i}^{\dagger} c_{j}^{\dagger} c_{i'} c_{j'}$, where $i$ ($j$) and $i'$ ($j'$) are chosen from $\{1,2\}$ ($\{3,4\}$) and $i \neq i'$ ($j \neq j'$). The $H_{Jh}$ term is ignored hereafter since it plays the same role as $H_{Jt}$ on the target qubit, but is two orders of magnitude smaller. 

Therefore, as shown in Fig.~\ref{config}(b), the control DQD influences the target DQD in two ways: (i) changing the energy detuning,
\begin{align}
& \Delta \varepsilon ^{(1,1)} = U_{23}+U_{24}-U_{13}-U_{14}, \notag \\
& \Delta \varepsilon ^{(0,2)} = 2U_{24}-2U_{14},
\label{Uij}
\end{align} 
and (ii) changing the tunnel coupling,
\begin{align}
\Delta t^{(1,1)} = J_{t3}+J_{t4}, ~
\Delta t^{(0,2)} = 2J_{t4}.
\label{Jt}
\end{align}
Here the superscripts $(1,1)$ and $(0,2)$ denote the charge states of the control DQD, which are determined by its spin state (triplet $T_{0}$ or singlet $S$) via Pauli blockade \cite{Petta.05,vanWeperen.11}. Taking Eqs.~(\ref{Uij}) and (\ref{Jt}) into account, the Hamiltonian of the target DQD is rewritten in a matrix form
\begin{align}
H_{T,\rm {eff}}^{\alpha}=\left( \! \begin{array}{cccc}
-\varepsilon_{\rm{eff}}^{\alpha}+U & -\sqrt{2}t_{\rm{eff}}^{\alpha} & J_{p} & 0\\
-\sqrt{2}t_{\rm{eff}}^{\alpha} & V+J_{e} & -\sqrt{2}t_{\rm{eff}}^{\alpha} & 0\\
J_{p} & -\sqrt{2}t_{\rm{eff}}^{\alpha} & \varepsilon_{\rm{eff}}^{\alpha}+U & 0\\
0 & 0 & 0 & V-J_{e}\end{array} \! \right),
\label{HTeff}
\end{align}
where the basis sets are $\{C_{1,\uparrow}^{\dagger}C_{1,\downarrow}^{\dagger} \left| 0 \right \rangle $, $[(C_{1,\uparrow}^{\dagger}C_{2,\downarrow}^{\dagger} + C_{2,\uparrow}^{\dagger}C_{1,\downarrow}^{\dagger}) $ $/ \sqrt{2}] \left| 0 \right \rangle $, $C_{2,\uparrow}^{\dagger}C_{2,\downarrow}^{\dagger} \left| 0 \right \rangle $, $[(C_{1,\uparrow}^{\dagger}C_{2,\downarrow}^{\dagger} - C_{2,\uparrow}^{\dagger}C_{1,\downarrow}^{\dagger})$ $/ \sqrt{2}] \left| 0 \right \rangle \}$, $U$ is the on-site Coulomb interaction, $V$ is the inter-site Coulomb interaction, $\varepsilon_{\rm{eff}}^{\alpha}=\varepsilon_{T}+\Delta \varepsilon^{\alpha}$,  $t_{\rm{eff}}^{\alpha}=t+J_{t}+\Delta t^{\alpha}$, $\alpha=(1,1)$ or $(0,2)$, and the definition of $J_{e}$, $J_{p}$, $J_{t}$, and $\varepsilon_{T}$ are shown in footnote [\onlinecite{JeJpJt}]. Diagonalizing the above Hamiltonian gives the exchange energy $J^{\alpha}$, which is the energy difference between the two lowest eigenvalues (one singlet and the other triplet). As a result, the coherent precession of the target qubit is controlled by the charge state of the control DQD via $J^{\alpha}$ \cite{Petta.05,Burkard.99}.

Using the configuration interaction method, all the coupling parameters in Eqs.~(\ref{Uij})-(\ref{HTeff}) can be readily calculated for a given confinement potential using the lowest-energy Fock-Darwin states \cite{Burkard.99,Hu.00,Liqz.10,Yang.11,Wang.11}. In this paper, we adopt the quadratic confinement potential $V(x,y)={\rm min}$ $[{W(x_{1},y_{1})-\mu_{1}, \cdots, W(x_{4},y_{4})-\mu_{4}, 0}]$, where $W(x_{i},y_{i})$ $=m \omega_{0}^{2}$ $[(x-x_{i})^{2}+(y-y_{i})^{2}]/2$ represents the quantum dot centered at $(x_{i},y_{i})$. 
The numerical calculations are carried out for a GaAs system \cite{Hu.00,Burkard.99} with $\hbar \omega_{0}=3.96$ meV, $a=30$ nm, and $d=100$ nm [Fig.~\ref{config}(a)].

\begin{figure}
\includegraphics[width=0.9\columnwidth]{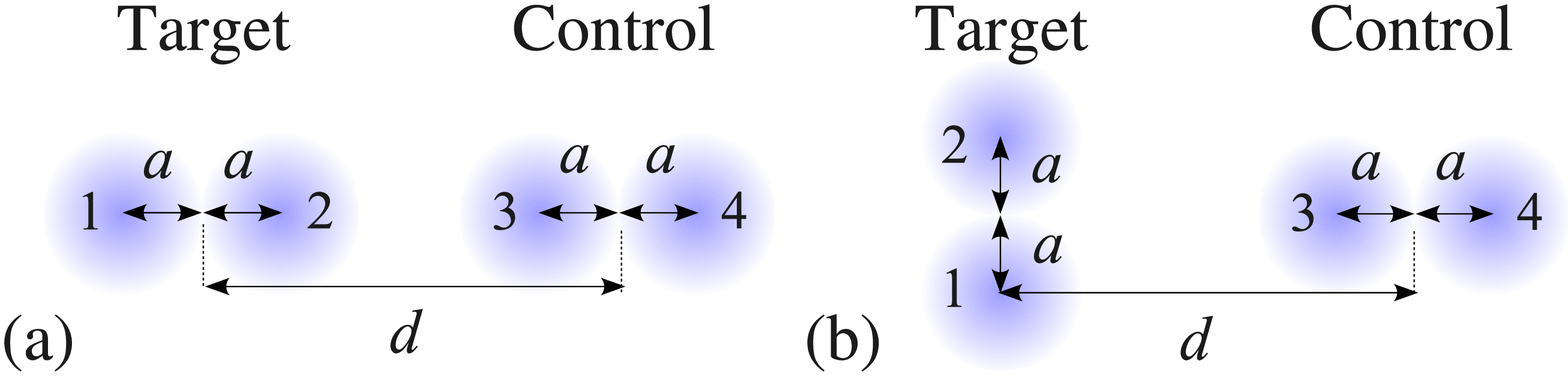}
\includegraphics[width=0.9\columnwidth]{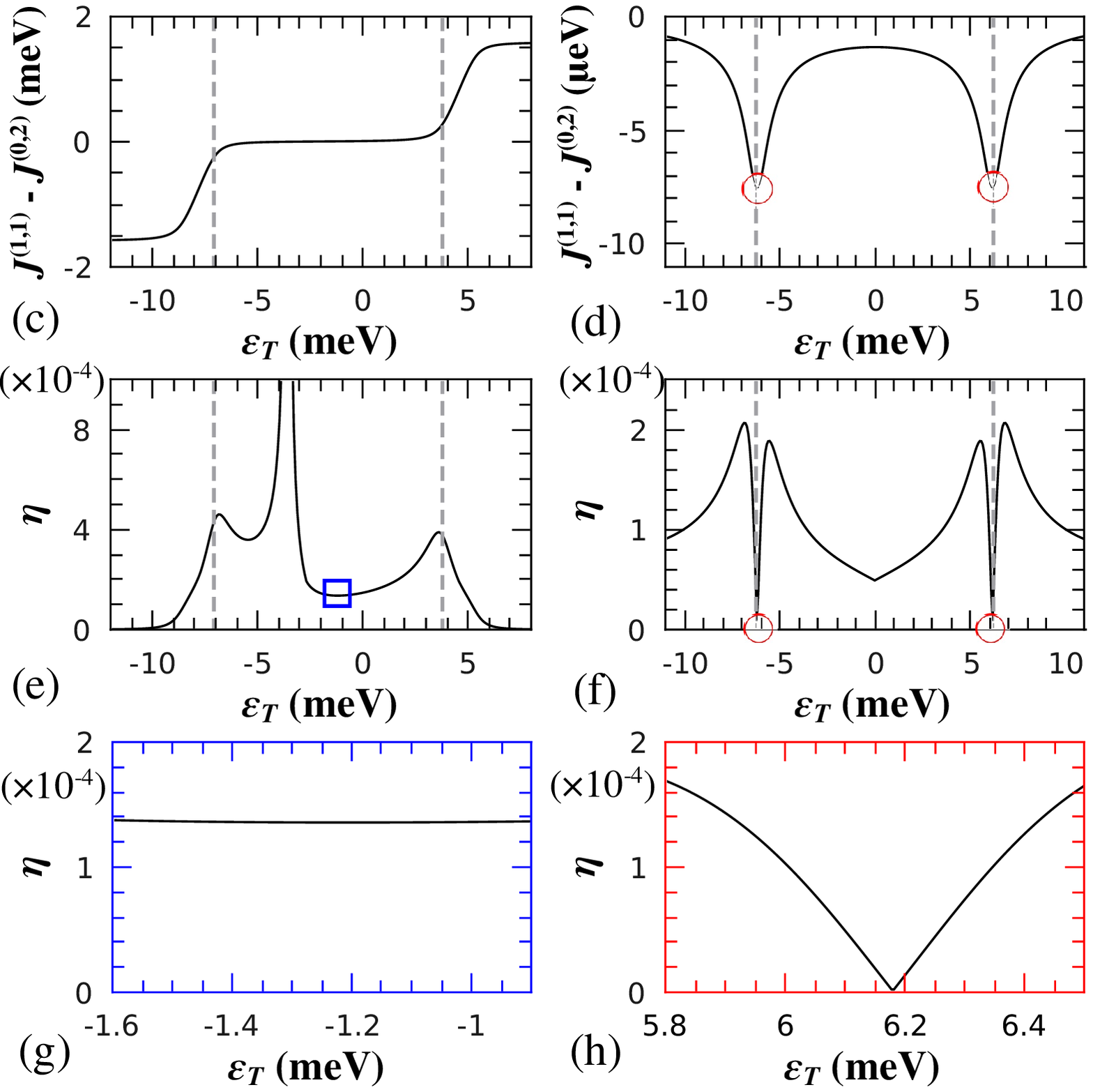}
\caption{(Color online) (a)-(b) Two typical configurations of the control and target DQDs. (c)-(d) Differences between the exchange energies with respect to the bias of the target DQD $\varepsilon_{T}$. (e)-(f) Relative errors of the controlled phase gate with respect to $\varepsilon_{T}$. (g)-(h) Close-ups of the low-noise spots. The panels on the left are for configuration (a), those on the right are for configuration (b).}
\label{case17}
\end{figure}

In Fig.~\ref{config}(c)-\ref{config}(d), the results of $\Delta \varepsilon^{(1,1)}-\Delta \varepsilon^{(0,2)}$ and $\Delta t^{(1,1)}-\Delta t^{(0,2)}$ are plotted as functions of $\theta_{C}$ and $\theta_{T}$. Their different patterns indicate that the ratio $R=[\Delta \varepsilon^{(1,1)}-\Delta \varepsilon^{(0,2)}]/[\Delta t^{(1,1)}-\Delta t^{(0,2)}]$ varies with the device geometry. Therefore, the geometric configuration of the system determines whether the energy detuning or the tunnel coupling plays a more important role. As an example, we consider two typical configurations shown in Fig.~\ref{case17}(a)-\ref{case17}(b). For configuration (a) with $\theta_{C}=\theta_{T}=0$, Fig.~\ref{config}(c)-\ref{config}(d) gives $R \approx 8$, which means that the energy detuning dominates the control process. However, for configuration (b) with $\theta_{C}=0$ and $\theta_{T}=\pi/2$, one finds $R=0$, indicating that the tunnel coupling plays the decisive role. As will be shown later, the ratio $R$ is crucial for reducing charge noise.

Next, we quantitatively investigate the gate error arising from charge noise. To facilitate the following discussion, we define a Bloch sphere for the target qubit, with $\left| S \right\rangle$ being the north pole and $\left| T_{0} \right\rangle$ being the south pole \cite{Petta.05,vanWeperen.11}. Following the experiments \cite{Petta.05}, the coherent manipulation of the target qubit consists of the following three steps: (i) preparing an initial state and adiabatically loading it into the $x$-$y$ plane of the Bloch sphere, (ii) rotating it about the $z$-axis through an angle $\theta^{\alpha}=J^{\alpha}\tau/\hbar$ during the precession time $\tau$, and (iii) adiabatically unloading it out of the $x$-$y$ plane and measuring the final state. We note that in step (ii), different control states result in different rotation angles $\theta^{\alpha}$. This difference [$\theta=\theta^{(1,1)}-\theta^{(0,2)}$] finally gives rise to a controlled $\theta$-phase gate.

In reality, unavoidable environmental charge noise affects the confinement potential of quantum dots, and therefore perturbs the exchange energy $J^{\alpha}$. For exchange errors $\delta J^{(1,1)}$ and $\delta J^{(0,2)}$, $\theta$ becomes $\theta+\delta \theta$, where $\delta \theta=[\delta J^{(1,1)}-\delta J^{(0,2)}]\tau/\hbar$. If we aim to perform a $\theta$-phase gate, the gate time should be $\tau=\hbar \theta/|J^{(1,1)}-J^{(0,2)}|$, and the relative error is given by
\begin{align}
\eta=| \delta \theta /\theta |=|\delta J^{(1,1)}-\delta J^{(0,2)}|/|J^{(1,1)}-J^{(0,2)}|.
\end{align}
Although there are many sources of charge noise \cite{Taylor.05,Taylor.07,Burkard.99,Romito.07,Borras.11} (background charge noise \cite{Burkard.99,Hu.06,Culcer.09,Nguyen.11}, gate noise \cite{Hu.00}, etc.), their effects on the target DQD are similar: raising or lowering the central barrier $V_{B}$ (thus changing in tunnel coupling $t$) and detuning the energy difference $\varepsilon_{T}$. Thus we have
\begin{align}
\eta=\frac{ \left | \frac{d[J^{(1,1)}-J^{(0,2)}]}{dt} \frac{dt}{dV_{B}} \delta V_{B}+\frac{d[J^{(1,1)}-J^{(0,2)}]}{d \varepsilon_{T}} \delta \varepsilon_{T} \right |} {|J^{(1,1)}-J^{(0,2)}|}.
\label{etaformula}
\end{align}
$\eta=0$ defines a sweet spot for the target qubit. To evaluate $\eta$, we make the reasonable assumption that $\delta V_{B}$ and $\delta \varepsilon_{T}$ are of the same magnitude and independent \cite{Jung.04}. For GaAs quantum dots the fluctuation is found to be $\delta V_{B} \approx \delta \varepsilon_{T} \approx 0.07-0.16 \mu$eV \cite{Jung.04,Hu.06,Culcer.09}. We choose $\delta V_{B}=\delta \varepsilon_{T}=0.2 \mu$eV in our calculation, which corresponds to the worst case scenario with stronger fluctuations. All the other terms in Eq. (\ref{etaformula}) can be readily calculated using the configuration interaction method \cite{Burkard.99,Hu.00,Liqz.10,Yang.11,Wang.11}.

So far we have shown how charge fluctuations affect the conditional operation. In the following, we will find optimal conditions, or ``sweet spots'', that suppress $\eta$. For practical purposes, it would be advantageous to keep $|J^{(1,1)}-J^{(0,2)}|$ appreciable at the same time. This turns out to be possible as we show below. In Fig.~\ref{case17}(c)-\ref{case17}(f), we plot calculated $J^{(1,1)}-J^{(0,2)}$ and $\eta$ as functions of $\varepsilon_{T}$ for the two typical configurations. For configuration (b), there exist two sweet spots near the crossovers of charge sectors. At the sweet spots, the relative error $\eta$ becomes zero, and $|J^{(1,1)}-J^{(0,2)}|$ reaches its maximum value. This means that if $\varepsilon_{T}$ can be fixed precisely at the sweet spots, we get a robust and relatively fast C-phase gate. In panel (f), $\eta$ seems to be sensitive to $\varepsilon_{T}$ in the vicinity of sweet spots. However, its close-up [panel (h)] shows that $\eta \leq 5 \times 10^{-5}$ as long as the error of $\varepsilon_{T}$ is no more than $\pm 0.08$ meV. Thus the low-noise gate can be performed with the state-of-the-art techniques. On the contrary, configuration (a) does not support any sweet spot in the $(1,1)_{T}$ charge sector. Only an optimal point with a relatively small $\eta$ is found at the blue square in panel (e). However, configuration (a) also offers some advantages. First, panels (c)-(d) show that the value of $|J^{(1,1)}-J^{(0,2)}|$ in case (a) is usually larger than the one in case (b), leading to a faster gate operation. Second, as can be seen from panel (g), working at the optimal point does not require a precise control of $\varepsilon_{T}$. In addition, the minimal possible $\eta$ ($\eta_{\rm{min}}=1.36 \times 10^{-4}$) is rather small, allowing in principle the possibility of fault-tolerant quantum computation. To summarize, we have shown the existence of sweet spots in case (a). We have also gone beyond the sweet spots and found an optimal point of operation for configuration (b). 

\begin{figure}
\includegraphics[width=0.48\columnwidth]{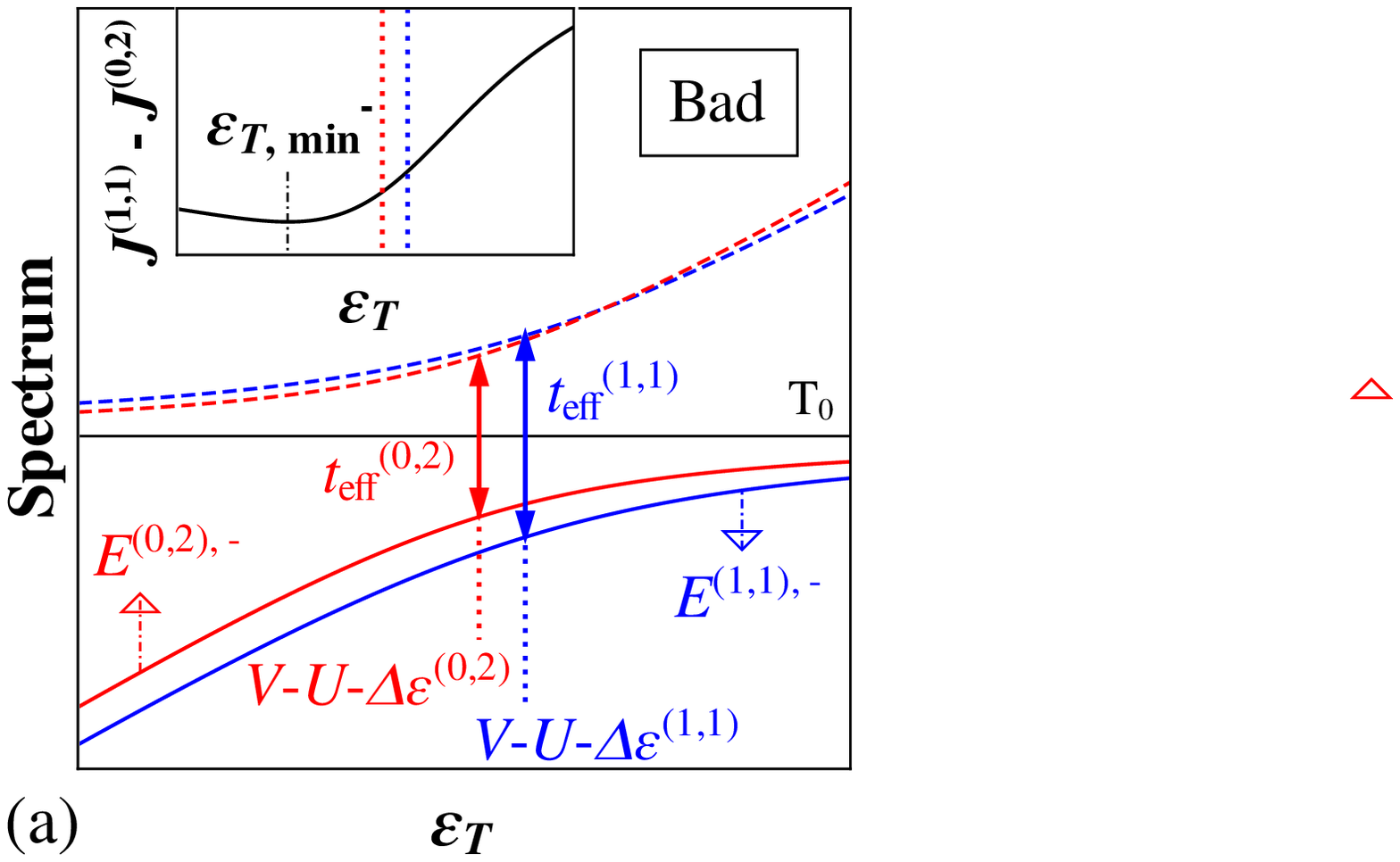}
\includegraphics[width=0.48\columnwidth]{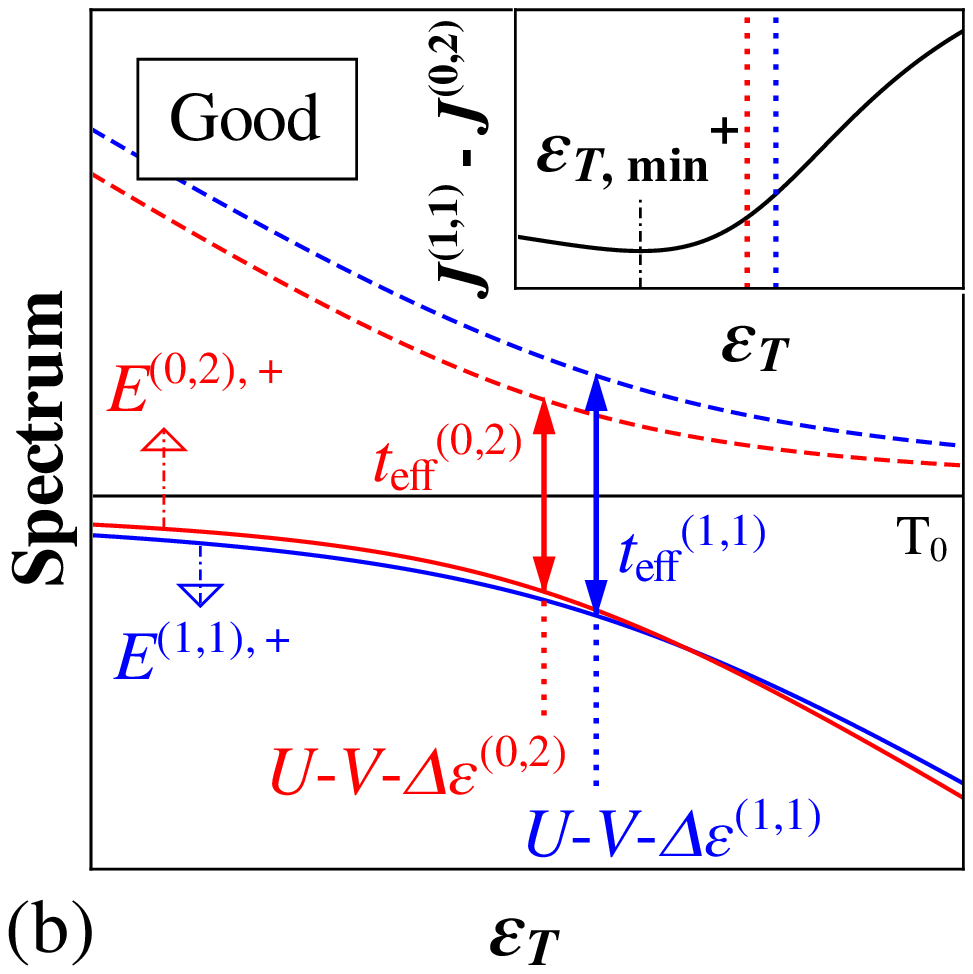}
\caption{(Color online) Low-energy spectra of the target DQD near the crossovers of the $(2,0)_{T}$ and $(1,1)_{T}$ charge sectors [panel (a)], and the $(1,1)_{T}$ and $(0,2)_{T}$ charge sectors [panel (b)]. The superscripts denote the charge states of the control DQD. Insets: The corresponding $J^{(1,1)}-J^{(0,2)}$ with respect to $\varepsilon_{T}$. Two sweet spots are found at $\varepsilon_{T}=\varepsilon_{T,\rm{min}}^{\pm}$. The dashed lines denote the boundaries between different charge sectors. A sweet spot in $(1,1)_{T}$ charge sector is good for reducing dephasing error.}
\label{perturb}
\end{figure}

Now we turn to the general configurations and provide analytic expressions for the sweet spots. The low-energy spectrum of the target DQD is shown in Fig.~\ref{perturb}(a)-\ref{perturb}(b). We note that near the crossovers of different charge sectors, the singlet states can be described approximately by a two-level anti-crossing model
\begin{align}
H_{T}^{' \alpha,\pm} & =\lambda_{\pm}^{\alpha}+\left(\begin{array}{cc}
\pm \lambda_{\pm}^{\alpha} & -\sqrt{2}t_{\rm{eff}}^{\alpha}\\
-\sqrt{2}t_{\rm{eff}}^{\alpha} & \mp \lambda_{\pm}^{\alpha} \end{array}\right),
\label{HTperturb}
\end{align}
where $\lambda_{\pm}^{\alpha}=(U-V \mp \varepsilon_{\rm{eff}}^{\alpha})/2$, the ``+'' superscript denotes the $(2,0)_{T}$ and $(1,1)_{T}$ crossover [see panel (a)], and the ``-'' superscript denotes the $(1,1)_{T}$ and $(0,2)_{T}$ crossover [see panel (b)]. Here we have neglected the small $J_{e}$ and $J_{p}$ terms for simplicity. Diagonalizing $H_{T}^{' \alpha,\pm}$ gives the energy of the lowest singlet state $E^{\alpha,\pm}=\lambda_{\pm}^{\alpha}$ $-\sqrt{2(t_{\rm{eff}}^{\alpha})^{2}+(\lambda_{\pm}^{\alpha})^{2}}$. On the other hand, as shown by the horizontal lines in Fig.~\ref{perturb}, the lowest triplet state has a constant energy. Since the exchange energy $J^{\alpha}$ is the gap between the lowest singlet and the triplet state, we obtain $[J^{(1,1)}-J^{(0,2)}]^{\pm}=E^{(0,2),\pm}-E^{(1,1),\pm}$. According to Eq.~(\ref{etaformula}), we can find the sweet spots by solving $d[J^{(1,1)}-J^{(0,2)}]^{\pm}/d \varepsilon_{T}=0$ and $d[J^{(1,1)}-J^{(0,2)}]^{\pm}/dt=0$ simultaneously. The sweet spots are given by
\begin{align}
\varepsilon_{T,\rm{min}}^{\pm} = \pm & (U-V)-\Delta \varepsilon^{(1,1)}+[t+J_{t}+\Delta t^{(1,1)}]R
\label{ETmin}
\end{align}
where $R=[\Delta \varepsilon^{(1,1)}-\Delta \varepsilon^{(0,2)}]/[\Delta t^{(1,1)}-\Delta t^{(0,2)}]$. However, there are some caveats here. First, Eq.~(\ref{HTperturb}) is a good approximation only when the sweet spots are close to the crossovers $\pm (U-V)-\Delta \varepsilon^{\alpha}$ (dashed lines in Fig.~\ref{perturb}), i.e., the absolute value of $R$ is small [see Eq.~(\ref{ETmin})]. Second, we want to find the sweet spots in the $(1,1)_{T}$ charge sector, i.e., $V-U-\min[\Delta \varepsilon^{(1,1)},\Delta \varepsilon^{(0,2)}] \leq \varepsilon_{T,\rm{min}} \leq U-V-\max[\Delta \varepsilon^{(1,1)},\Delta \varepsilon^{(0,2)}]$, because a high double-occupation probability would result in a large dephasing error \cite{Petta.05,vanWeperen.11}. In general, as can be seen from Eq.~(\ref{ETmin}) and the insets of Fig.~\ref{perturb}, only one sweet spot meets this requirement, given by the condition that $R<0$ ($R>0$), $\varepsilon_{T,\rm{min}}^{+}$ ($\varepsilon_{T,\rm{min}}^{-}$) is inside the $(1,1)_{T}$ regime. 

\begin{figure}
\begin{center}
\includegraphics[width=0.45\columnwidth]{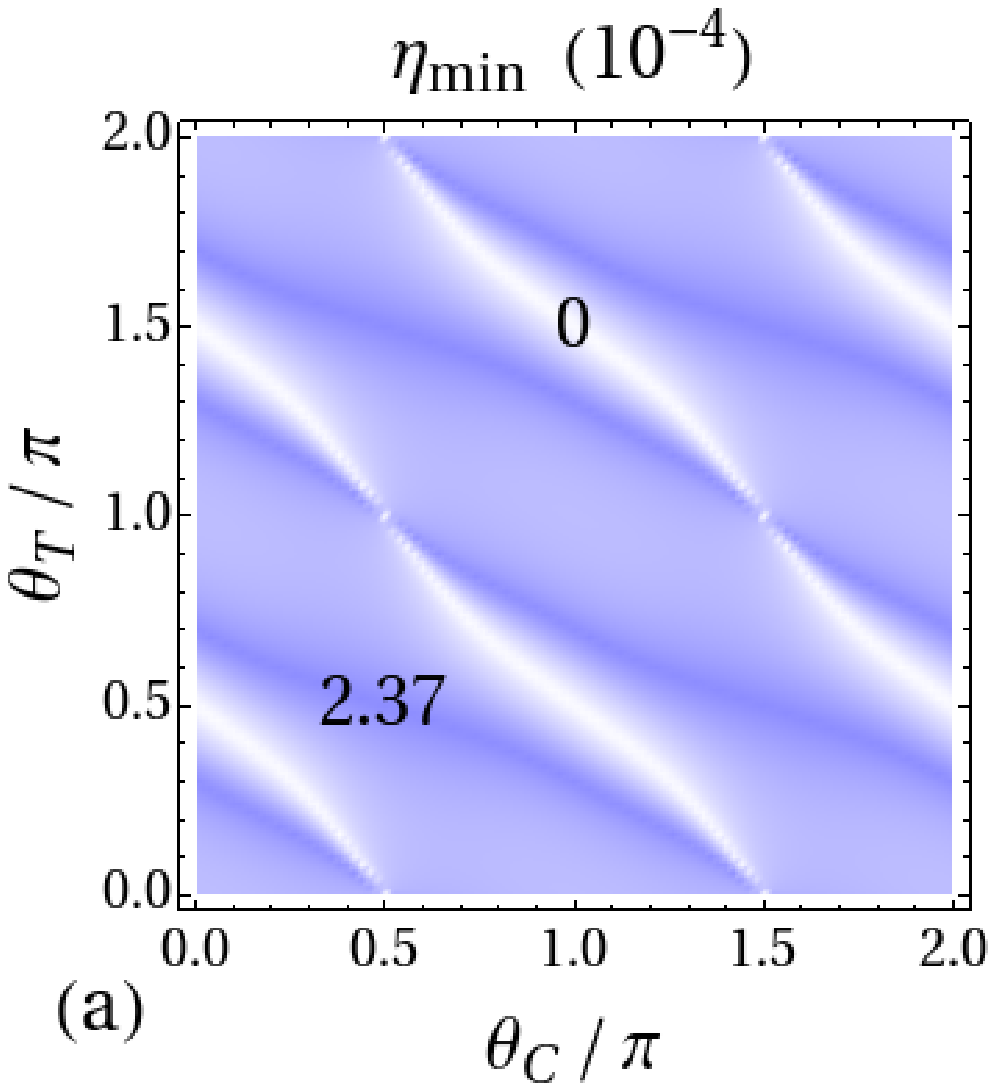}
\includegraphics[width=0.45\columnwidth]{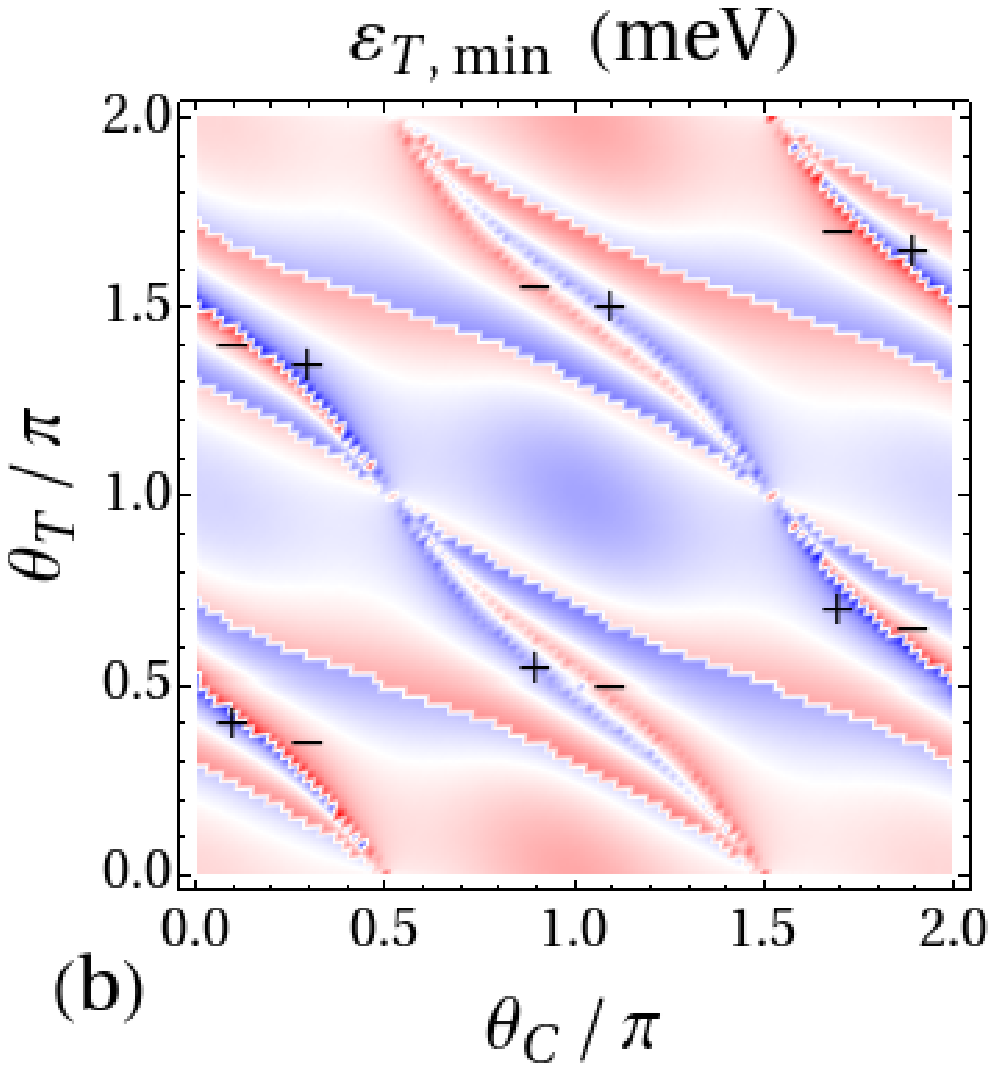}
\includegraphics[width=0.06\columnwidth]{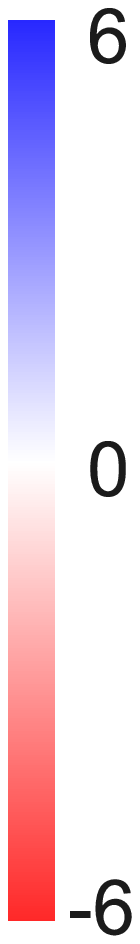}
\end{center}
\caption{(Color online) (a) In the $(1,1)_{T}$ charge sector, the minimal possible relative error $\eta_{\rm{min}}$ as a function of $\theta_{C}$ and $\theta_{T}$. (b) The optimal detuning of the target DQD with respect to $\theta_{C}$ and $\theta_{T}$.}
\label{noise}
\end{figure}

The above results are further verified by exact numerical calculations. As shown in Fig.~\ref{noise}(a), the minimal possible relative error in the $(1,1)_{T}$ charge sector is plotted as a function of $\theta_{C}$ and $\theta_{T}$. Comparing the white regions with Fig.~\ref{config}(c), one finds that the system is immune to charge noise when $\Delta \varepsilon^{(1,1)}-\Delta \varepsilon^{(0,2)} \approx 0$ (i.e., $R \approx 0$) and $\varepsilon_{T}=\varepsilon_{T,\rm{min}}$. In contrast, from the dark blue regions and Fig.~\ref{config}(c), we see that the system is always sensitive to charge noise when $\Delta t^{(1,1)}-\Delta t^{(0,2)} \approx 0$ (i.e., $R \rightarrow \infty$). Moreover, there are some light blue regions in between, where the system has an optimal point as the one shown in Fig.~\ref{case17}(a). In Fig.~\ref{noise}(b), $\varepsilon_{T,\rm{min}}$ of the sweet spots is also in excellent agreement with the analytical approximation in Eq.~(\ref{ETmin}). Therefore, the ratio $R$ and the device geometry together determine the sweet spots and their applicability in coherent qubit manipulations.

In summary, we have studied the coherent multiqubit operations in the coupled DQD system with a general geometry emphasizing how to reduce charge noise. We have developed a microscopic model to fully understand the mechanism of the conditional operation. We have shown that the exchange energy of the target qubit is affected by the control qubit through two channels: energy detuning and tunnel coupling. In particular, the device geometry determines which one plays a dominant role. Using the configuration interaction method, we have evaluated the relative error of the C-phase gate arising from charge noise. We have demonstrated the existence of noise-immune sweet spots in some optimal configurations, where the tunnel coupling serves as the main control channel. On the contrary, one always sees large charge fluctuations in the $(1,1)_{T}$ charge sector if the tunnel coupling makes no contribution. We have further developed a two-level anti-crossing model which analytically describes the conditions for sweet spots. In addition, we have found some optimal points in the parameter space, where the gate is fast and insensitive to charge noise. Our work should guide future experimental efforts to carry out coherent multiqubit operations in double-quantum-dot structures.

We thank J. P. Kestner and M. Cheng for helpful discussions. This work was supported by IARPA and LPS.


\begin{thebibliography}{[00]}
\bibitem{JeJpJt}
The Hamiltonian of the isolated target DQD is $H_{T}=H_{\varepsilon}+H_{t}+H_{U}+H_{J}$, where the chemical potential part is $H_{\varepsilon}=\sum_{\sigma} -(\varepsilon_{T}/2) n_{1 \sigma} +(\varepsilon_{T}/2) n_{2 \sigma}$, and the hopping terms are $H_{t}=\sum_{\sigma} (-t c_{1 \sigma}^{\dagger} c_{2 \sigma} +{\rm H.c.})$. The Coulomb repulsion terms are $H_{U}=U n_{1 \uparrow} n_{1 \downarrow} +U n_{2 \uparrow} n_{2 \downarrow}+V(n_{1 \uparrow} n_{2 \downarrow}+n_{1 \downarrow} n_{2 \uparrow}) + (V-J_{e}) (n_{1 \uparrow} n_{2 \uparrow}+n_{1 \downarrow} n_{2 \downarrow})$, and $H_{J}=-J_{e} c_{1 \downarrow}^{\dagger} c_{2 \uparrow}^{\dagger} c_{2 \downarrow} c_{1 \uparrow}-J_{p} c_{2 \uparrow}^{\dagger} c_{2 \downarrow}^{\dagger} c_{1 \uparrow} c_{1 \downarrow} -\sum_{i \sigma} J_{t} n_{i \sigma} c_{1 \overline{\sigma}}^{\dagger} c_{2 \overline{\sigma}} +{\rm H.c.}$, including the spin-exchange ($J_{e}$), pair-hopping ($J_{p}$) and occupation-modulated hopping terms ($J_{t}$). The Hamiltonian of the isolated control DQD ($H_{C}$) has a similar form as $H_{T}$, with $1 \rightarrow 3$, $2 \rightarrow 4$, and $\varepsilon_{T} \rightarrow \varepsilon_{C}$.
\end{thebibliography}
\end{document}